\documentclass[final,3p,times,twocolumn]{elsarticle}
\usepackage{graphicx}
\usepackage{amsmath,amssymb}
\journal{Astroparticle Physics}

\newcommand{\lp}{\left(}
\newcommand{\rp}{\right)}
\newcommand{\lb}{\left[}
\newcommand{\rb}{\right]}

\newcommand{\ba}{\begin{eqnarray}}
\newcommand{\ea}{\end{eqnarray}}
\newcommand{\be}{\begin{equation}}
\newcommand{\ee}{\end{equation}}

\newcommand{\ka}{\kappa}

\newcommand{\R}{\mathcal{R}}

\newcommand{\OA}{\Omega_A}

\begin{document}

\begin{frontmatter}

\title{Probing hybrid modified gravity by stellar motion around
Galactic Centre}

\author[a]{D. Borka}
\ead{dusborka@vinca.rs}
\author[b,c,d]{S. Capozziello}
\author[e]{P. Jovanovi\'{c}}
\author[a]{V. Borka Jovanovi\'{c}}
\address[a]{Atomic Physics Laboratory (040), Vin\v{c}a
Institute of Nuclear Sciences, University of Belgrade, P.O. Box 522,
11001 Belgrade, Serbia}
\address[b]{Dipartimento di  Fisica, Universit\`{a} di Napoli
"Federico II",
 Compl. Univ. di Monte S. Angelo, Edificio G, Via Cinthia, I-80126,
Napoli, Italy}
\address[c]{Istituto Nazionale di  Fisica Nucleare (INFN) Sez.
di
Napoli, Compl. Univ. di Monte S. Angelo,
Edificio G, Via Cinthia, I-80126, Napoli, Italy}
\address[d]{Gran Sasso Science Institute (INFN),  Viale F.
Crispi, 7, I-67100,
L'Aquila, Italy.}
\address[e]{Astronomical Observatory, Volgina 7, P.O. Box 74, 11060
Belgrade, Serbia}

\begin{abstract}
We consider possible signatures for the so called {\it hybrid
gravity} within the Galactic Central Parsec. This modified theory of
gravity consists of a superposition of the metric Einstein-Hilbert
Lagrangian with an $f(R)$ term constructed {\it \`{a} la Palatini}
and can be easily reduced to an equivalent scalar-tensor theory.
Such an approach is introduced in order to cure the shortcomings
related to $f(R)$ gravity, in general formulated either in metric or
in metric-affine frameworks. Hybrid gravity allows to disentangle
the further gravitational degrees of freedom with respect to those
of standard General Relativity. The present analysis is based on the
S2 star orbital precession around the massive compact dark object at
the Galactic Centre where the simulated orbits in hybrid modified
gravity are compared with astronomical observations. These 
simulations result with constraints on the range of hybrid gravity
interaction parameter $\phi_0$, showing that in the case of S2 star
it is between -0.0009 and -0.0002. At the same time, we are also able
to obtain the constraints on the effective mass parameter $m_{\phi}$,
and found that it is between -0.0034 and -0.0025 AU$^{-1}$ for S2
star. Furthermore, the hybrid gravity potential induces precession of
S2 star orbit in the same direction as General Relativity. In
previous papers, we considered other types of extended gravities,
like metric power law $f(R)\propto R^n$ gravity, inducing Yukawa and
Sanders-like gravitational potentials, but it seems that hybrid
gravity is the best among these models to explain different
gravitational phenomena at different astronomical scales.
\end{abstract}


\end{frontmatter}

\section{Introduction}
\label{sec:Sec1}

The existence of different anomalous astrophysical and cosmological
phenomena like the cosmic acceleration, the dynamics of galaxies
and gas in clusters of galaxies, the galactic rotation curves, etc.
recently boosted the growth of several long-range modifications
of the usual laws of gravitation. These mentioned phenomena did not
find satisfactory explanations in terms of the standard
Newton-Einstein gravitational physics, unless exotic and still
undetected forms of matter-energy are postulated: dark matter and
dark energy. A recent approach is to try to explain these phenomena
without using new material ingredients like dark matter and dark
energy, but using well-motivated generalization and extensions of
General Relativity (GR). Several alternative gravity theories have
been proposed (see e.g.
\cite{fisc99,clif12,capo11a,capo11b,cope04,soti10,noji11} for
reviews), such as: MOND \cite{milg83}, scalar-tensor
\cite{bran61,moff05,moff06,bisw12}, conformal \cite{behn02,barb06},
Yukawa-like corrected gravity theories \cite{fisc92,card11,stab,
stab1}, theories of "massive gravity"
\cite{ruba08,babi10,pitt07,babi09,rham11,yun13,babi13}. Alternative
approaches to Newtonian gravity in the framework of the weak field
limit \cite{clif05} of fourth-order gravity theory have been
proposed and constraints on these theories have been discussed
\cite{capo06,capo07,zakh06,zakh07,frig07,nuci07,bork12,capo09a,
iori10,bork13,capo14,zakh14}.

The philosophy is to search for alternative form of gravity, i.e. of
the Einstein-Hilbert theory, so that such modifications could
naturally explain some astrophysical and cosmological phenomena
without invoking the presence of new material ingredients that,
at the present state of the art, seem hard to be detected. Besides,
this extended approach can be connected to effective theories that
emerge both from the quantization on curved spacetimes and from
several unification schemes \cite{clif12,capo11a,capo11b}.

The simplest extension of the Einstein-Hilbert action is based on
straightforward generalizations of the Einstein theory where the
gravitational action (the Einstein-Hilbert action) is assumed to be
linear in the Ricci curvature scalar $R$. If this action consists in
modifying the geometric part considering a generic function $f(R)$,
we get so called $f(R)$ gravity which was firstly proposed in 1970
by Buchdahl \cite{buch70}. Generally, the most serious problem of
$f(R)$ theories is that these theories cannot easily pass the
standard Solar System tests \cite{chib03,olmo05}. However, there
exists some classes of them that can solve this problem 
\cite{noji03}. It can be shown that $f(R)$ theories, in principle,
could explain the evolution of the Universe, from a matter dominated
early epoch up to the present, late-time self accelerating phase.
Several debates are open in this perspective
\cite{amen07a,gohe09,capo06b,amen07b} but the crucial point is that
suitable self-consistent model can be achieved. $f(R)$ theories have
also been studied in the Palatini approach, where the metric and the
connection are regarded as independent fields \cite{olmo}. Metric
and Palatini  approaches are certainly equivalent in the context of
GR, i.e., in the case of the linear Einstein-Hilbert action. This is
not so for extended gravities. The Palatini variational approach
leads to second order differential field equations, while the
resulting field equations in the metric approach are fourth order
coupled differential equations. These differences also extend to the
observational aspects.

A novel approach, that  consists of adding to the metric
Einstein-Hilbert Lagrangian an $f(R)$ term constructed within the
framework of the Palatini formalism, was recently proposed
\cite{hark12,capo13,capo13a}. The aim of this formulation is twofold:
from one side, one wants to describe the extra gravitational budget
in metric-affine formalism, from the other side, one wants to cure
the shortcomings emerging in $f(R)$ gravity both in metric
and Palatini formulations. In particular, hybrid gravity allows to
disentangle the metric and the geodesic structures pointing out that
further degrees of freedom coming from $f(R)$ can be recast as an
auxiliary scalar field. In such a case, problems related to the
Brans-Dicke-like representation of $f(R)$ gravity in terms of
scalar-tensor theory (the so called O'Hanlon transformation) are
immediately avoided (see \cite{capo13a} for details and the
discussion in Sec. 2). Due to this feature, the scalar-tensor
representation of hybrid gravity results preferable with respect to
other scalar-tensor representations of gravitational interaction. As
byproducts, the appearance of ghosts is avoided and the correct weak
field limit of $f(R)$ gravity with respect to GR is recovered.
Furthermore, several issues related to the galactic dynamics, the
formulation of the virial theorem in alternative gravity, the dark
energy behavior seem to be better addressed than in $f(R)$ gravity
considered in both metric or Palatini formulations. In summary, the
hybrid metric-Palatini theory opens up new possibilities to approach,
in the same theoretical framework, the problems of both dark energy
and dark matter disentangling the extra degrees of freedom of
gravitational field with respect to GR. For a brief review on the
hybrid metric-Palatini theory, we refer the reader to \cite{capo13b}.

In this perspective, star dynamics around the Galactic Centre could
be a useful test bed to probe the effective gravitational potentials
coming from the theory. In particular, S-stars are the young bright
stars which move around the centre of our Galaxy
\cite{ghez00,scho02,gill09a,gill09b,ghez08,genz10} where the compact
radio source Sgr A$^\ast$ is located. For more details about S2 star
see references \cite{gill12,genz10}. There are some observational
indications that the orbits of some of them, like S2, could deviate
from the Keplerian case \cite{gill09a,meye12}, but the current
astrometric limit is not sufficient to unambiguously confirm such a 
claim \cite{bork13,frit10}.

Here we study a possible application of hybrid modified gravity
within Galactic Central Parsec, in order to explain the observed
precession of orbits of S-stars. This paper is a continuation of
previous studies where we considered different extended gravities,
such as  power law $f(R)$ gravity \cite{bork12,zakh14}, $f(R,\phi)$
gravity implying Yukawa and Sanders-like gravitational potentials in
the weak field limit \cite{bork13, capo14}. Results obtained using
hybrid gravity point out that, very likely, such a theory is the
best candidate among those considered to explain (within the same
picture) different gravitational phenomena at different astronomical
scales.

More details about hybrid gravity you can find in
\cite{olmo,hark12,capo13a,capo13b}. It is shown in
\cite{capo13a,capo13b} that this type of modified gravity is
coherently addressing the Solar System issues, and motivations for
addressing them are discussed in details in \cite{capo13b}.

The modified theory of gravity needs to be constrained at different
scales: at laboratory distances, at Solar system, at galaxies, at 
galactic clusters and at cosmological scales. Obtaining constraints 
at any of these scales is a fundamental issue to select or rule out 
models. In particular, it is important to investigate gravity in the 
vicinity of very massive compact objects because the environment 
around these objects is drastically different from that in the Solar 
System framework. The S2 star orbit is a unique opportunity to test  
gravity at the sub-parsec scale of a few thousand AU. For example, 
gravity is relatively well constrained at short ranges (especially 
at sub-mm scale) by experimental tests, however for long ranges 
further tests are still needed (see Figures 9 and 10 from 
\cite{adel09} for different ranges). It is worth stressing that a 
phenomenological approach can be useful in this context. In 
particular, the motion of S2-star is a suitable tool to test 
alternative gravity. For the reasons that we will discuss in detail 
below, hybrid gravity is a reliable paradigm to describe 
gravitational interaction without considering dark energy and dark 
matter. Specifically, the massive compact object inside the 
Galactic Center is surrounded by a matter distribution and deviations 
of S2-star motion from the Keplerian orbit are observed in detail. 
These deviations can be triggered both by the masses of the 
surrounding bodies and by the strong field regime at the Galactic 
Center. This peculiar situation constitutes a formidable opportunity 
to test theories of gravity. However, it is important to stress that 
numerical results reported here by comparing models with 
astronomical observations, represent only upper bounds for the 
precession angle on the deviation from GR. More accurate studies will 
be necessary in future work to better constrain dynamics around the 
Galactic Center.

The present paper is organized as follows: in Sec. 2 we sketch the
theory of hybrid gravity. In Sec. 3 we describe our simulations of
stellar orbits in the gravitational potential derived in the weak
field limit of hybrid gravity and we describe the fitting procedure
to simulate orbits with respect to astrometric observations of S2
star. Results are presented in Sec. 4. Conclusions are drawn in Sec.
5.

\section{Hibrid metric-Palatini gravity}
\label{sec:Sec2}

In this Section, we present the basic formalism for the hybrid
metric-Palatini gravitational theory within the equivalent
scalar-tensor representation (we refer the reader to
\cite{capo13a,capo13b,koiv10,capo12} for more details). The $f(R)$
theories are the special limits of the one-parameter class of
theories where the scalar field depends solely on the stress energy
trace $T$ (Palatini version) or solely on the Ricci curvature $R$
(metric version). Here, we consider a one-parameter class of
scalar-tensor theories where the scalar field is given as an
algebraic function of the trace of the matter fields and the scalar
curvature \cite{koiv10}:

\begin{equation}
\label{st_action} S = \int d^D x \sqrt{-g}\left[\frac{1}{2}\phi R -
\frac{D-1}{2(D-2)\left(\Omega_A-\phi\right)}(\partial\phi)^2 -
V(\phi)\right].
\end{equation}
The theories can be parameterized by the constant $\Omega_A$. The
limiting values $\Omega_A=0$ and $\Omega_A \rightarrow \infty$
correspond to scalar-tensor theories with the Brans-Dicke parameter
$\omega=-(D-1)/(D-2)$ and $\omega=0$. These limits reduce to $f(R)$
gravity in the Palatini and the metric formalism, respectively. For
any finite value of $\Omega_A$, its value depends both on  matter and
curvature. In the limit $\Omega_A \rightarrow \infty$ the propagating
mode is given solely by the curvature, $\phi(R,T) \rightarrow
\phi(R)$, and in the limit $\Omega_A\rightarrow 0$ solely the matter
fields $\phi(R,T) \rightarrow \phi(T)$. In the general case, the
field equations are fourth order both in the matter and in the metric
derivatives as we will show below.

More specifically, the intermediate theory with $\Omega_A=1$  and
$D=4$, corresponds to the hybrid metric-Palatini gravity theory
proposed in \cite{hark12, capo13a}, where the action is given by
\begin{equation}
\label{action} S= \int d^4 x \sqrt{-g} \left[ R + f(\mathcal{R}) +
2\kappa^2
\mathcal{L}_m \right]\,.
\end{equation}
where $\kappa^2\equiv 8\pi G$, $R$ is the Einstein-Hilbert
term, $\R  \equiv g^{\mu\nu}\R_{\mu\nu} $ is the Palatini curvature
with the independent connection $\hat{\Gamma}^\alpha_{\mu\nu}$  as
\be \R \equiv  g^{\mu\nu} \R_{\mu\nu} \equiv g^{\mu\nu}\lp
\hat{\Gamma}^\alpha_{\mu\nu , \alpha}
       - \hat{\Gamma}^\alpha_{\mu\alpha , \nu} +
\hat{\Gamma}^\alpha_{\alpha\lambda}\hat{\Gamma}^\lambda_{\mu\nu} -
\hat{\Gamma}^\alpha_{\mu\lambda}\hat{\Gamma}^\lambda_{\alpha\nu}\rp\,
.\label{r_def}
\ee
The Palatini-Ricci tensor  $\R_{\mu\nu}$ is
\begin{equation}
\R_{\mu\nu} \equiv \hat{\Gamma}^\alpha_{\mu\nu ,\alpha} -
\hat{\Gamma}^\alpha_{\mu\alpha , \nu} +
\hat{\Gamma}^\alpha_{\alpha\lambda}\hat{\Gamma}^\lambda_{\mu\nu}
-\hat{\Gamma}^\alpha_{\mu\lambda}\hat{\Gamma}^\lambda_{\alpha\nu}\,.
\end{equation}
Varying the action given with respect to the metric, one obtains the
field equations
\be
\label{efe} G_{\mu\nu} +
F(\R)\R_{\mu\nu}-\frac{1}{2}f(\R)g_{\mu\nu} = \ka^2 T_{\mu\nu}\,,
\ee
where the matter stress-energy tensor is
 \be \label{memt}
 T_{\mu\nu} \equiv -\frac{2}{\sqrt{-g}} \frac{\delta
 (\sqrt{-g}\mathcal{L}_m)}{\delta(g^{\mu\nu})}.
 \ee
The independent connection is compatible with the metric
$F(\R)g_{\mu\nu}$, conformal to $g_{\mu\nu}$, with the conformal
factor given by $F(\R) \equiv df(\R)/d\R$. This fact gives
\begin{equation}
\begin{array}{l}
\label{ricci} \R_{\mu\nu} = R_{\mu\nu} +
\frac{3}{2}\frac{1}{F^2(\R)}F(\R)_{,\mu}F(\R)_{,\nu} \\
~~~~~ - \frac{1}{F(\R)}\nabla_\mu F(\R)_{,\nu}
- \frac{1}{2}\frac{1}{F(\R)}g_{\mu\nu}\nabla_\alpha
\nabla^\alpha F(\R)\,.
\end{array}
\end{equation}

The Palatini curvature $\R$ is obtained from the trace of the field
equations (\ref{efe}), which is \be \label{trace} F(\R)\R -2f(\R)=
\ka^2 T +  R \equiv X\,.
\ee
$\R$ can be algebraically expressed in terms of $X$ if $f(\R)$ is
analytic. In other words, the variable $X$ measures how the theory
deviates from GR trace equation $R=-\ka^2 T$.

We can express the field equations (\ref{efe}) in terms of the metric
and $X$ as
\ba \label{efex}
G_{\mu\nu} & = &
\frac{1}{2}f(X)g_{\mu\nu}- F(X)R_{\mu\nu}  +  F'(X)
\nabla_{\mu}X_{,\nu} \nonumber \\
& + & \frac{1}{2}\lb F'(X)\nabla_\alpha \nabla^\alpha X + F''(X)\lp
\partial X\rp^2 \rb
g_{\mu\nu} \nonumber \\
& + & \lb F''(X)-\frac{3}{2}\frac{\lp
F'(X)\rp^2}{F(X)}\rb X_{,\mu}X_{,\nu} + \ka^2 T_{\mu\nu}\,, \ea
being $(\partial X)^2=X_{,\mu}X^{,\mu}$.
The trace of the field equations is now \ba \label{trace2}
F'(X)\nabla_\alpha \nabla^\alpha X + \lb
F''(X)-\frac{1}{2}\frac{\lp F'(X)\rp^2}{F(X)}\rb \lp \partial
X\rp^2 \nonumber \\
+ \frac{1}{3}\left[ X + 2f(X)-F(X)R\right]= 0 \,,
\ea  while the relation between the metric  scalar curvature  $R$
and the Palatini scalar curvature  $\R$ is \be \label{ricciscalar}
\R(X) = R+\frac{3}{2}\lb
\lp\frac{F'(X)}{F(X)}\rp^2-2\frac{\nabla_\alpha \nabla^\alpha
F(X)}{F(X)}\rb\,, \ee which can be obtained by contracting
Eq.~(\ref{ricci}). As for pure metric and Palatini cases
\cite{capo11b}, the action (\ref{action}) for the hybrid
metric-Palatini theory can be recast into a scalar-tensor theory by
an auxiliary field $A$ such that
\begin{equation} \label{eq:S_scalar0}
S= \frac{1}{2\kappa^2}\int d^4 x \sqrt{-g} \left[R +
f(A)+f_A(\R-A)\right] +S_m \ ,
\end{equation}
where $f_A\equiv df/dA$ and $S_m$ is the matter action. Rearranging
the terms and defining $\phi\equiv f_A$, $V(\phi)=A f_A-f(A)$, Eq.
(\ref{eq:S_scalar0}) becomes
\begin{equation} \label{eq:S_scalar1}
S= \frac{1}{2\kappa^2}\int d^4 x \sqrt{-g} \left[R +
\phi\R-V(\phi)\right] +S_m \ .
\end{equation}
The variation of this action with respect to the metric, the scalar
$\phi$ and the connection leads to the field equations
\begin{eqnarray}
R_{\mu\nu}+\phi
\R_{\mu\nu}-\frac{1}{2}\left(R+\phi\R-V\right)g_{\mu\nu}&=&\kappa^2
T_{\mu\nu}\,,
\label{eq:var-gab}\\
\R-V_\phi&=&0 \label{eq:var-phi} \,, \\
\hat{\nabla}_\alpha\left(\sqrt{-g}\phi g^{\mu\nu}\right)&=&0 \,,
\label{eq:connection}\
\end{eqnarray}
respectively.
The solution of Eq.~(\ref{eq:connection}) implies that the
independent connection is the Levi-Civita connection of a metric
$h_{\mu\nu}=\phi g_{\mu\nu}$, that is we are dealing with a
bi-metric theory and $\R_{\mu\nu}$ and $R_{\mu\nu}$ are related by
\begin{equation} \label{eq:conformal_Rmn}
\R_{\mu\nu}=R_{\mu\nu}+\frac{3}{2\phi^2}\partial_\mu \phi
\partial_\nu \phi-\frac{1}{\phi}\left(\nabla_\mu
\nabla_\nu \phi+\frac{1}{2}g_{\mu\nu}\nabla_\alpha \nabla^\alpha
\phi\right) \ ,
\end{equation}
which can be used in the action (\ref{eq:S_scalar1}) to obtain the
following scalar-tensor representation
\begin{equation} \label{eq:S_scalar2}
S= \frac{1}{2\kappa^2}\int d^4 x \sqrt{-g} \left[ (1+\phi)R
+\frac{3}{2\phi}\partial_\mu \phi \partial^\mu \phi
-V(\phi)\right] +S_m \ .
\end{equation}
We have to stress that, by the substitution $\phi \rightarrow
-(\kappa\phi)^2/6$, the action (\ref{eq:S_scalar2}) reduces to the
case of a conformally coupled scalar field with a self-interaction
potential. This redefinition makes the kinetic term in the action
(\ref{eq:S_scalar2}) the standard one, and the action itself becomes
that of a massive scalar-field conformally coupled to the Einstein
gravity. Of course, it is not the Brans-Dicke gravity where the
scalar field is massless.

As discussed above, in the limit $\OA\rightarrow 0$, the theory
(\ref{eq:S_scalar2}) becomes the Palatini-$f(\R)$ gravity, and in the
limit $\OA\rightarrow \infty$ it is the metric $f(R)$ gravity. Apart
from these singular cases, any theory with a finite $\OA$ is in the
''hybrid'' regime, which from this point of view provides a unique
interpolation between the two a priori completely distinct classes
of gravity theories.

Using Eq.~(\ref{eq:conformal_Rmn}) and Eq.~(\ref{eq:var-phi}) in
Eq.~(\ref{eq:var-gab}), the metric field equations are
\begin{eqnarray}
(1+\phi) R_{\mu\nu} & = &
\kappa^2\left(T_{\mu\nu}-\frac{1}{2}g_{\mu\nu}
T\right) \nonumber \\
&& + \frac{1}{2}g_{\mu\nu}\left(V+\nabla_\alpha
\nabla^\alpha \phi\right) \nonumber \\
&& +\nabla_\mu\nabla_\nu\phi-\frac{3}{2\phi}
\partial_\mu \phi
\partial_\nu \phi \ \label{eq:evol-gab}\,,
\end{eqnarray}
and then the spacetime curvature is sourced by both matter and
scalar field. The scalar field equation can be manipulated in two
different ways that illustrate how this theory is related with the
$w=0$ and $w=-3/2$ cases, which corresponds to the metric and
Palatini scalar-tensor representations of $f(R)$-gravity
\cite{capo11b} respectively. Considering the trace of
Eq.~(\ref{eq:var-gab}) with $g^{\mu\nu}$, we find
$-R-\phi\R+2V=\kappa^2T$, and using Eq.~(\ref{eq:var-phi}), it is
\begin{equation}\label{eq:phi(X)}
2V-\phi V_\phi=\kappa^2T+ R \ .
\end{equation}
Similarly as in the Palatini case ($w=-3/2$), this equation says
that the field $\phi$ can be expressed as an algebraic function of
the scalar $X\equiv \kappa^2T+ R$, i.e., $\phi=\phi(X)$. In the pure
Palatini case, however, $\phi$ is just a function of $T$. Therefore
the right-hand side of Eq.~(\ref{eq:evol-gab}) contains matter terms
associated with the trace $T$, its derivatives, and also the
curvature $R$ and its derivatives. In other words, this theory can
be seen as a higher-derivative theory in both matter and  metric
fields. However, such an interpretation can be avoided if $R$ is
replaced in Eq. (\ref{eq:phi(X)}) with the relation
\begin{equation}
R=\R+\frac{3}{\phi}\nabla_\mu \nabla^\mu
\phi-\frac{3}{2\phi^2}\partial_\mu \phi \partial^\mu \phi
\end{equation}
together with $\R=V_\phi$. One then finds that the scalar field
dynamics is given by a second-order equation that becomes, for
$\OA=1$,
\begin{equation}\label{eq:evol-phi}
-\nabla_\mu \nabla^\mu \phi+\frac{1}{2\phi}\partial_\mu \phi
\partial^\mu
\phi+\frac{\phi[2V-(1+\phi)V_\phi]} {3}=\frac{\phi\kappa^2}{3}T\,,
\end{equation}
which is a Klein-Gordon equation. This result shows that, unlike in
the Palatini case ($w=-3/2$), the scalar field is dynamical. In this
sense, the theory is not affected by the microscopic instabilities
that arise in Palatini models (see \cite{olmo} for details).

\section{The weak field limit and the fitting procedure}

In the weak field limit and far from the sources, the scalar field
behaves as $\phi(r) \approx \phi_0 + ( 2G\phi_0 M /3r) e^{-m_\phi
r}$; the effective mass is defined as
\begin{equation}
\label{mass}
m_\phi^2 \equiv \left.
(2V-V_{\phi}-\phi(1+\phi)V_{\phi\phi})/3\right| _{\phi=\phi_0}\,,
\end{equation}
where $\phi_0$ is the amplitude of the background value of $\phi$.
Furthermore $V$, $V_\phi$ and $V_{\phi\phi}$ are respectively the
potential and its first and the second derivatives with respect to
$\phi$. The metric perturbations yield
\begin{eqnarray}
\label{pippo}
h_{00}^{(2)}(r) &=& \frac{2G_{\rm eff} M}{r} +\frac{V_0}{1+\phi_0}
\frac{r^2}{6}\,, \nonumber \\
h_{ij}^{(2)}(r) &=& \left(\frac{2\gamma
G_{\rm eff} M}{r}-\frac{V_0}{1+\phi_0}\frac{r^2}
{6}\right)\delta_{ij} \label{cor3}\ ,
\end{eqnarray}
where $V_0$ is the minimum of the potential $V$. The effective
Newton constant $G_{\rm eff}$ and the post-Newtonian parameter
$\gamma$ are defined as
\begin{eqnarray}
\label{pippo1}
G_{\rm eff} &\equiv&
\frac{G}{1+\phi_0}\left[1-\left(\phi_0/3\right)e^{-m_\phi
r}\right]\,, \nonumber \\
\gamma &\equiv& \frac{1+\left(\phi_0/3\right)e^{-m_\phi
r}}{1-\left(\phi_0/3\right)e^{- m_\phi r}} \,.
\end{eqnarray}

The coupling of the scalar field to the local system depends on
$\phi_0$. If $\phi_0 \ll 1$, then $G_{\rm eff}\approx G$ and 
$\gamma\approx 1$ regardless of the value of $m_\phi^2$. This is in 
contrast with the result obtained in the metric version of $f(R)$ 
theories. For sufficiently small $\phi_0$, this modified theory 
allows to pass the Solar System tests, even if the scalar field is 
very light \cite{capo13b}. According to these considerations, the 
leading parameters are $m_\phi$ and $\phi_0$. Their value give both 
an estimation of the deviation with respect to GR and how the 
affine contribution (i.e. the Palatini term) is relevant with respect 
to the metric $f(R)$ gravity. Constraining both of them by 
observations gives immediately information on the hybrid gravity. 
Starting from the above results, the modified gravitational potential 
can be written in  the form:

\begin{eqnarray}
\label{pot}
\Phi \left( r \right)=
-\frac{G}{1+\phi_0}\left[1-\left(\phi_0/3\right)e^{-m_\phi
r}\right] M/r.
\end{eqnarray}
An important remark is necessary at this point. We have not chosen
the form of $V(\phi)$ since the only requirement is that the scalar
field potential is an analytic function of $\phi$. In such a case,
the effective mass (\ref{mass}) can be always defined. Clearly, the
aim is to derive specific forms of the potential starting from the
observations. This means a sort of "inverse scattering procedure"
by which the $V(\phi)$ potential can be reconstructed from the
observed values of the parameters $M$, $\phi_0$, $m_\phi$ and
$\gamma$.

To this end, let us use eq. (\ref{pot}) to simulate orbits of S2
star in the hybrid modified gravity potential and then we compare
the obtained results with the set of S2 star observations obtained
by the New Technology Telescope/Very Large Telescope (NTT/VLT). The
simulated orbits of S2 star are obtained by numerical integration of
equations of motion where the hybrid  gravitational potential is
adopted, i.e.
\begin{equation}
\mathbf{\dot{r}}=\mathbf{v},\hspace*{0.5cm}
\mu\mathbf{\ddot{r}}=-\triangledown\Phi\left(
\mathbf{r}\right),
\label{2body}
\end{equation}

\noindent where $\mu$ is the reduced mass in the two-body
problem. In that way we obtained the simulated orbit of S2 star
around Galactic Centre in the weak field approximation of hybrid
gravity where eqs. (\ref{pippo}) and (\ref{pippo1}) stand. Taking
into account that $\gamma$ = $\gamma$($\phi_0$, $m_{\phi}$), the
considered weak field solution depends on the following three
parameters: $M$, $\phi_0$, and $m_{\phi}$. Mass $M$ of the central
object can be obtained independently using different observational
techniques, such as e.g. virial analysis of the ionized gas in the
central parsec \cite{lacy82} (yielding $M = 3 \times 10^6 M_{sun}$),
$M-\sigma$ (mass - bulge velocity dispersion) relationship for the
Milky Way \cite{trem02} (yielding $M = 9.4 \times 10^6 M_{sun}$) or
from Keplerian orbits of S-stars \cite{gill09b} (yielding $M = 4.3
\times 10^6 M_{sun}$). Since our goal was not to make a new estimate
of mass $M$ using hybrid gravity, but instead to study the possible
deviations from Keplerian orbit of S2 star which could indicate
signatures for hybrid gravity on these scales, we adopted the last of
three previously mentioned estimates for mass of the central object
($M = 4.3 \times 10^6 M_{sun}$), as well as the distance to the S2
star given by \cite{gill09a} ($d_\star$ = 8.3 kpc), and constrained
only the remaining two free parameters ($\phi_0$, $m_{\phi}$).
Parameter $\phi_0$ is dimensionless, while $m_{\phi}$ is given in
AU$^{-1}$ (AU being astronomical unit), so that $m_{\phi}^{-1}$
represents a scaling parameter for gravity interaction. Non-zero
values of these two parameters, if obtained, would indicate a
potential deviation from GR.

In order to obtain the constraints on $\phi_0$ and $m_{\phi}$, these
two parameters were varied. For each their combination the simulated 
coordinates $x$ and $y$ and velocity components $v_x$ and $v_y$ of 
S2 star were calculated. Calculations were performed for each 
observational epoch and then compared with its corresponding observed 
positions and velocities. $\chi^2$ between the observed and 
calculated coordinates of S2 star is minimized using LMDIF1 routine 
from MINPACK-1 Fortran 77 library which solves the nonlinear least 
squares problems by a modification of Marquardt-Levenberg algorithm 
\cite{more80} (for more details on fitting procedures see 
\cite{bork13}).

\begin{figure*}
\centering
\includegraphics[width=0.45\textwidth]{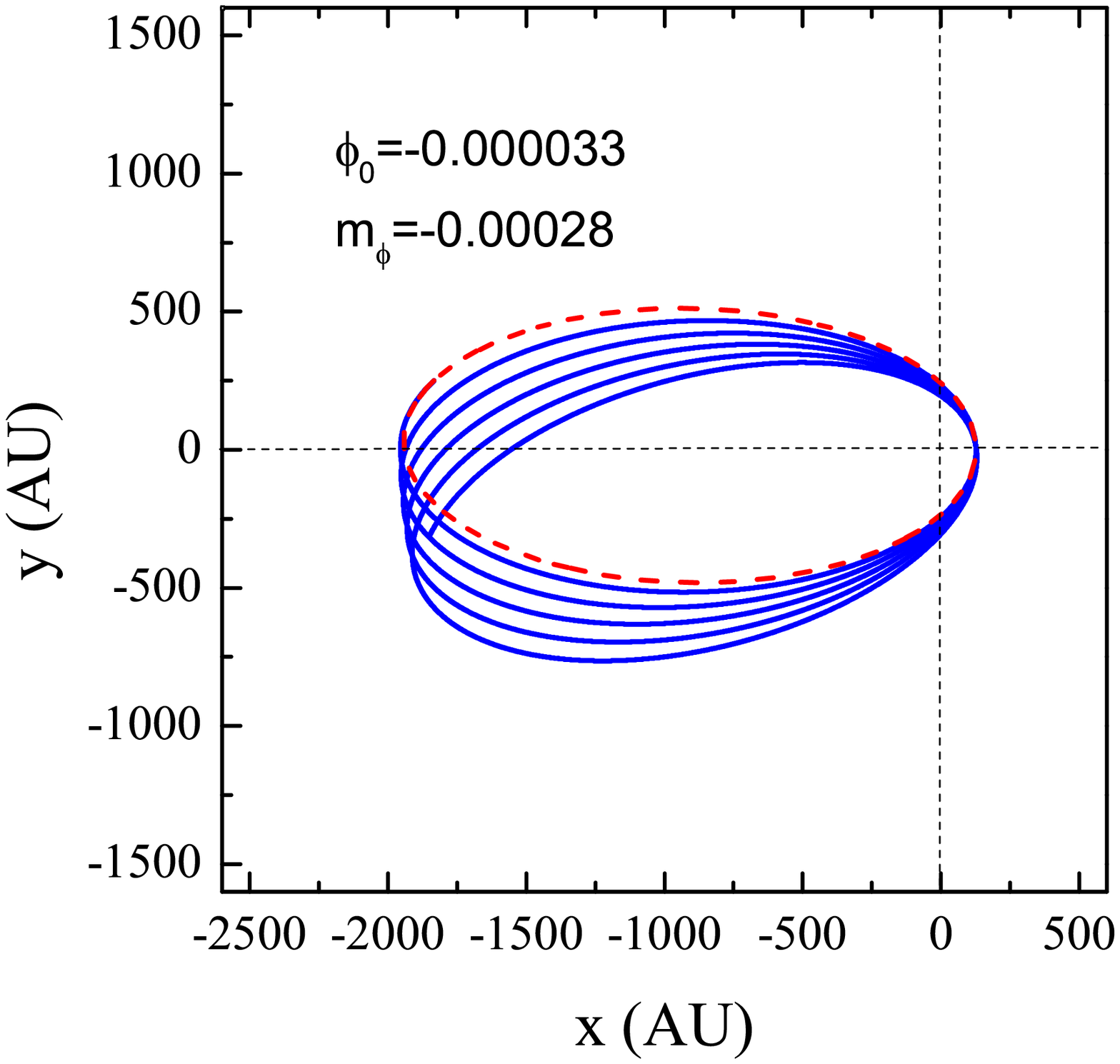}
\hspace*{0.8cm}
\includegraphics[width=0.45\textwidth]{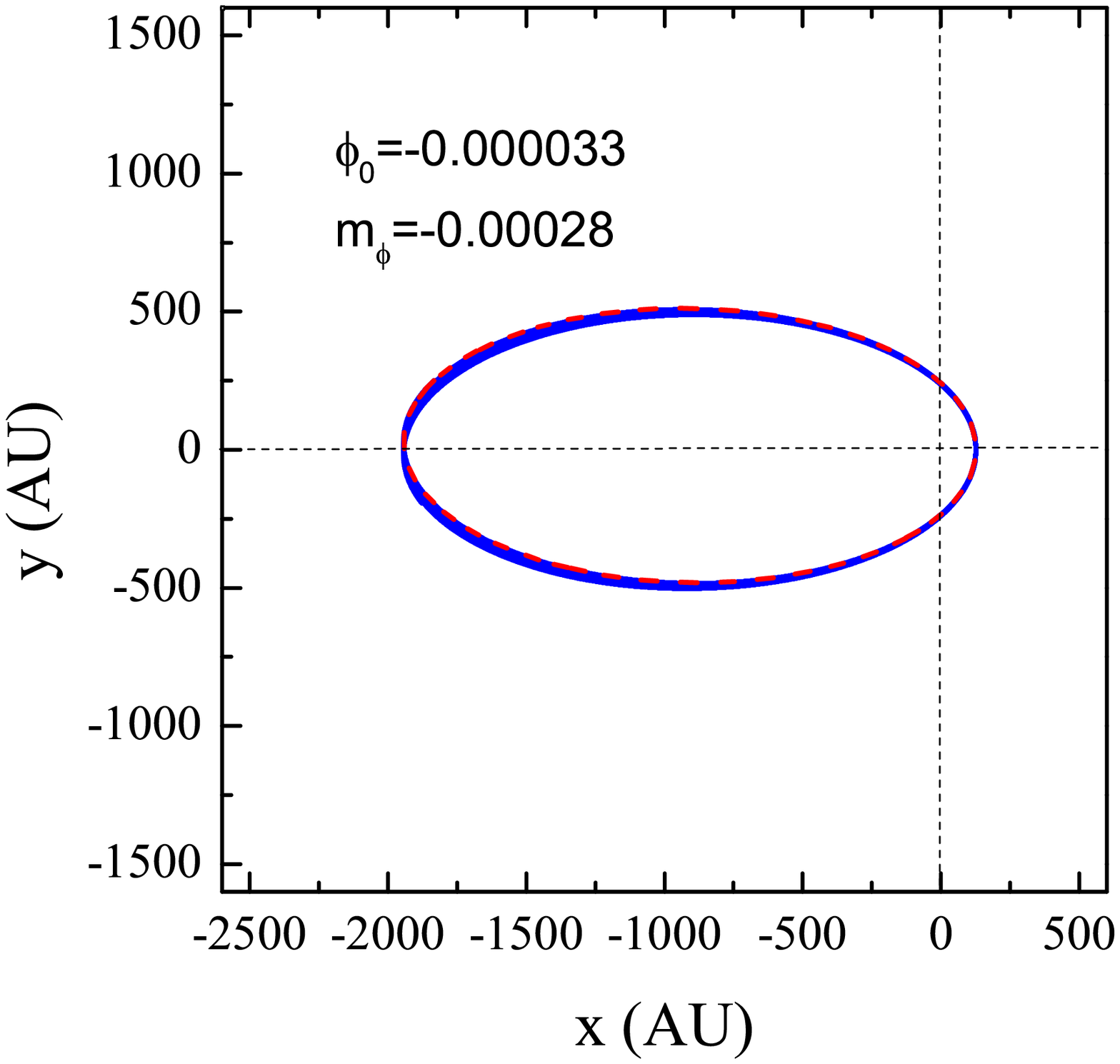}
\caption{(Color online) Comparisons between the orbit of S2
star in Newtonian gravity (red dashed line) and hybrid gravity
during 5 orbital periods (blue solid line) for (left panel) $\phi_0$
= -0.00033 and $m_\phi$ = -0.0028, and for (right panel) $\phi_0$ =
-0.000033 and $m_\phi$ = -0.00028.}
\label{fig01}
\end{figure*}

\begin{figure*}
\centering
\includegraphics[width=0.45\textwidth]{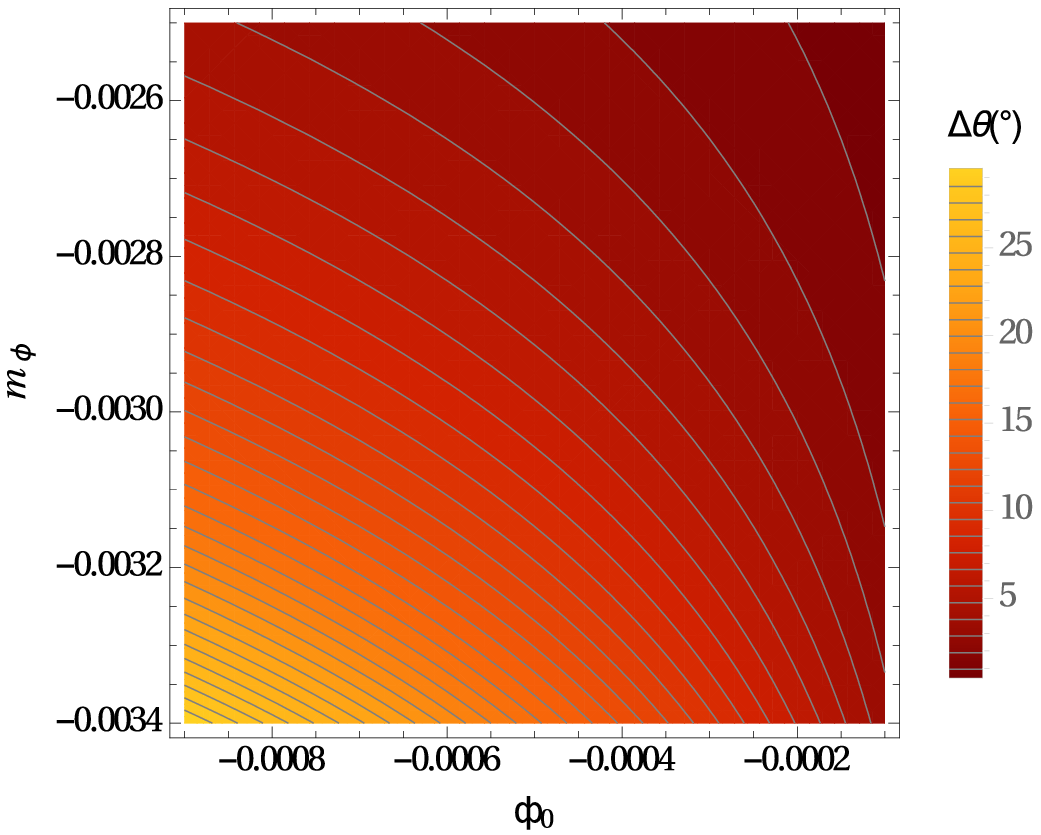}
\hspace*{0.8cm}
\includegraphics[width=0.45\textwidth]{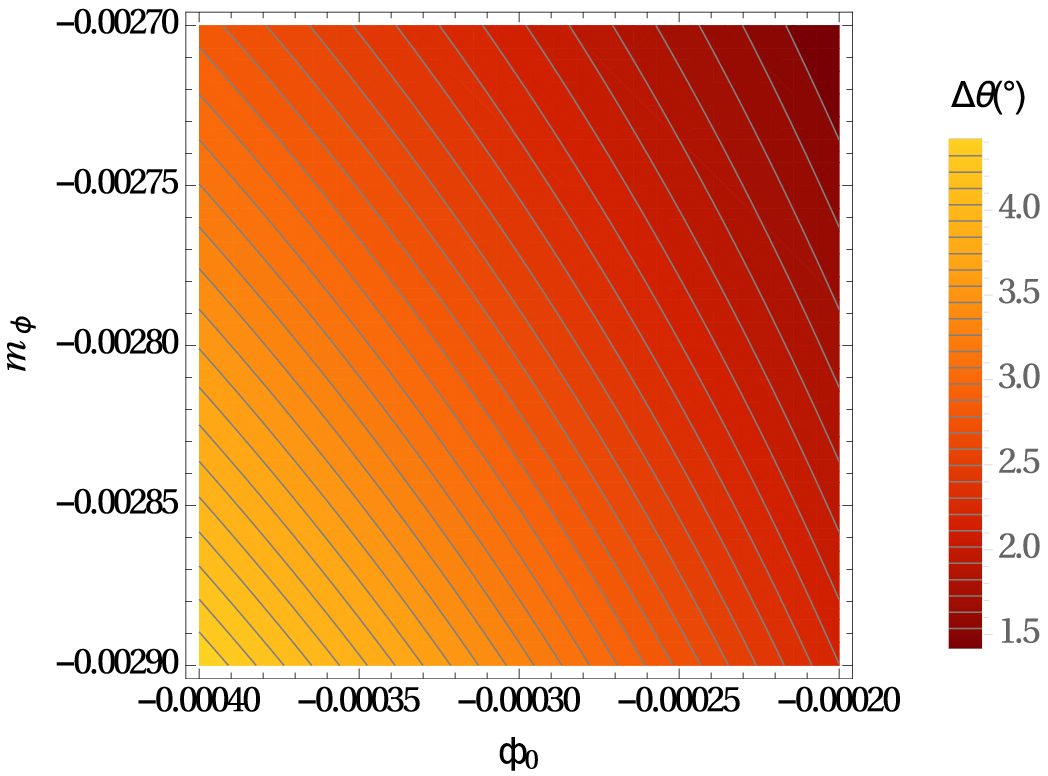}
\caption{(Color online) The precession per orbital period for
$\phi_0$ in the range $[-0.0009, -0.0002]$ and $m_\phi$ in
$[-0.0034, -0.0025]$ (left panel), and  $\phi_0$ in the range
$[-0.0004, -0.0002]$ and $m_\phi$ in $[-0.0029, -0.0027]$
(right panel) in the case of hybrid modified gravity potential. With
a decreasing value of angle of precession colors are darker.}
\label{fig02}
\end{figure*}

\begin{figure*}
\centering
\includegraphics[width=0.45\textwidth]{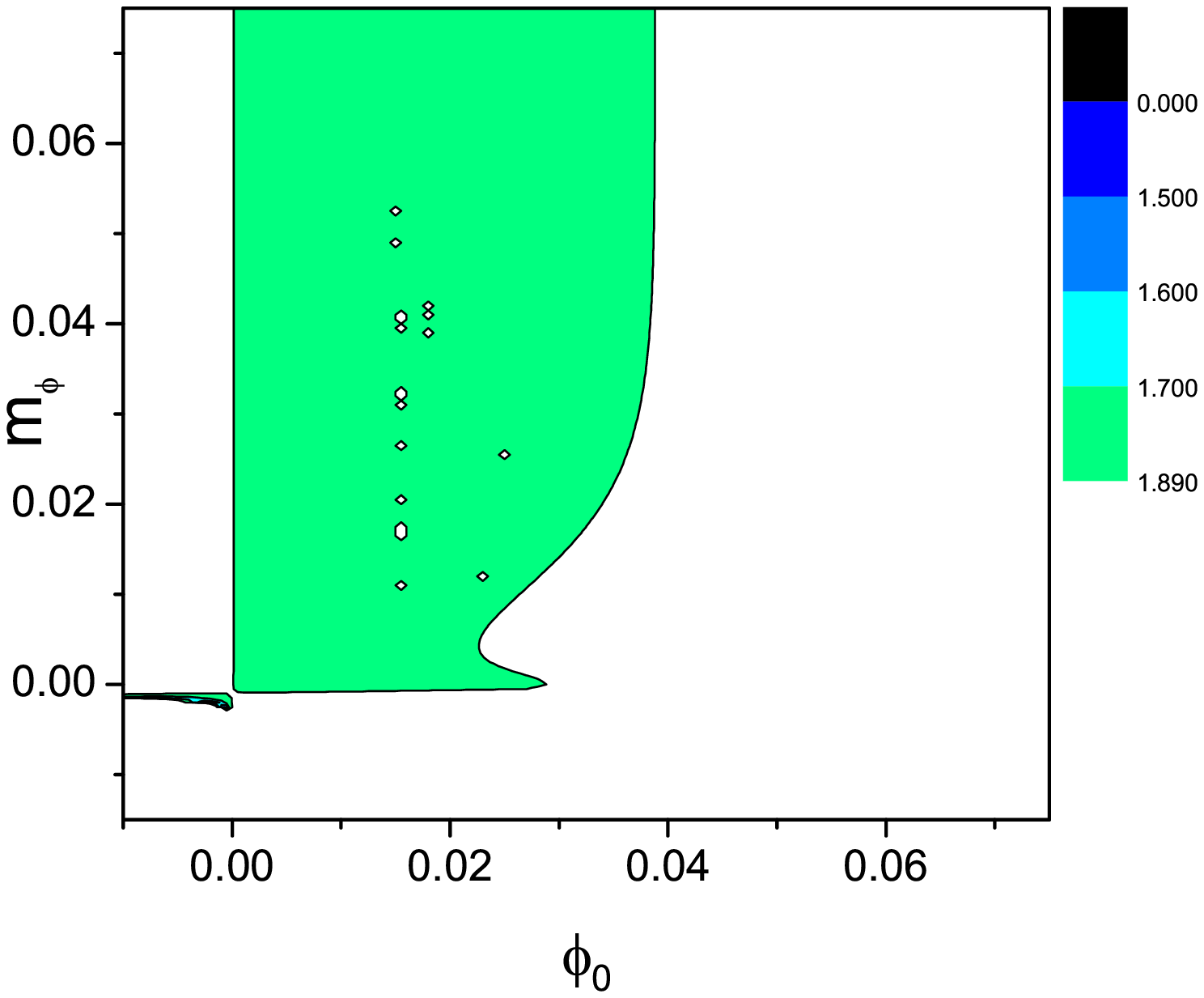}
\hspace*{1cm}
\includegraphics[width=0.45\textwidth]{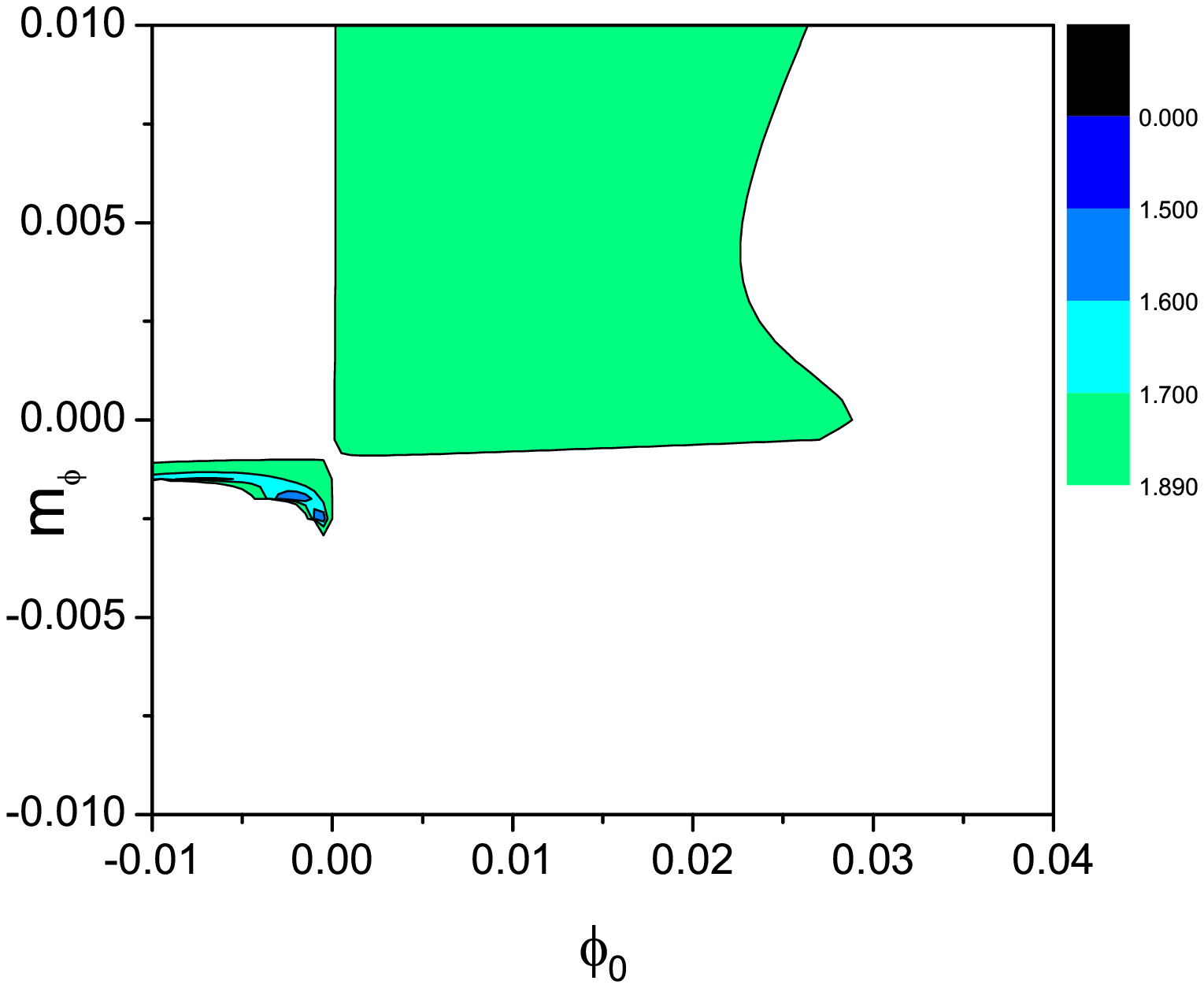}
\caption{(Color online) The maps of the reduced $\chi^{2}$ over the
$\phi_0 - m_\phi$ parameter space for all simulated orbits of S2
star which give at least the same or better fits than the Keplerian
orbits. With a decreasing value of $\chi^{2}$ (better fit) colors
ingrey scale are darker. A few contours are presented for specific
values of reduced $\chi^{2}$ given in the figure's legend.}
\label{fig03}
\end{figure*}

\begin{figure*}
\centering
\includegraphics[width=0.45\textwidth]{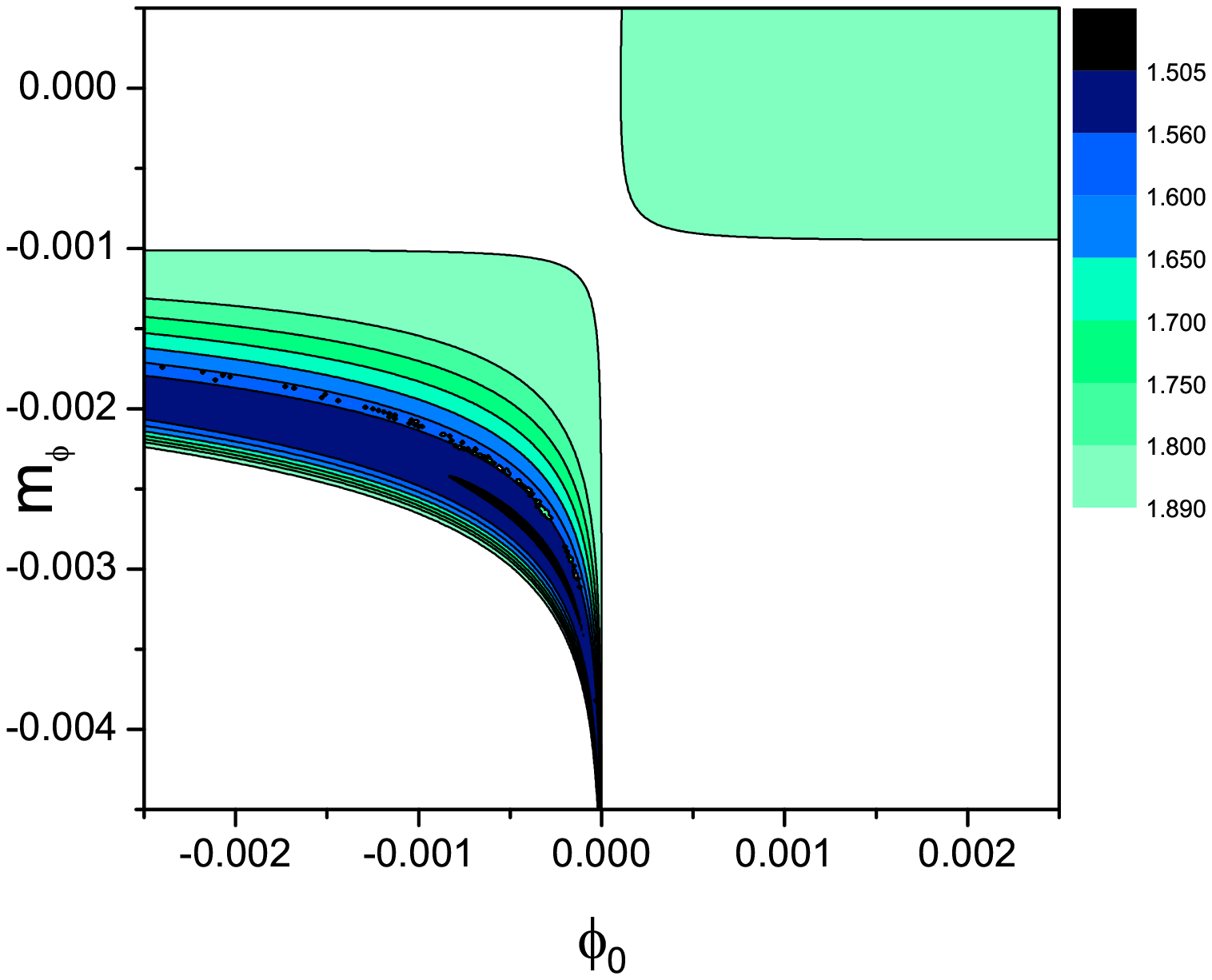}
\hspace*{1cm}
\includegraphics[width=0.45\textwidth]{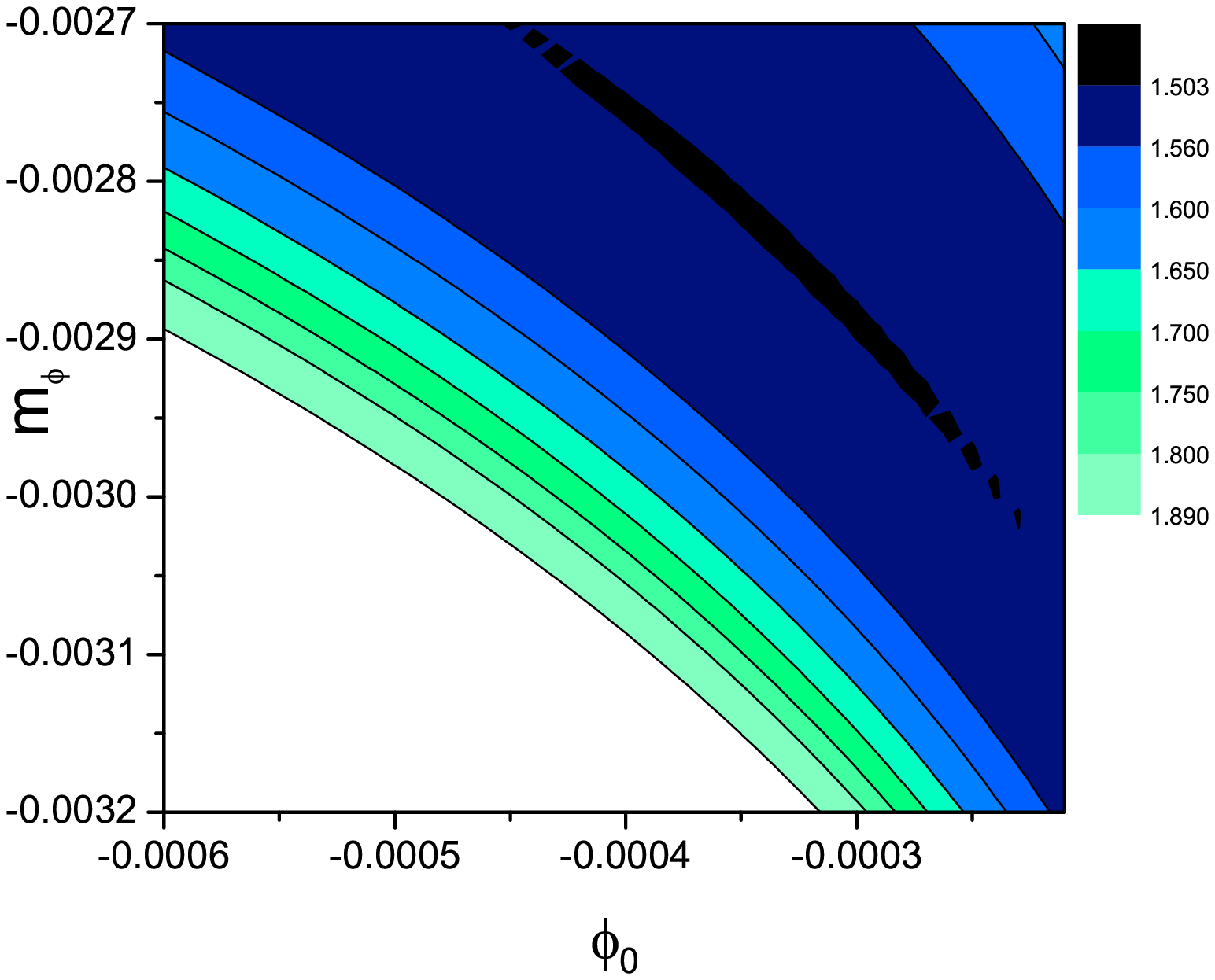}
\caption{(Color online) The same as in Fig. \ref{fig03}, but for the
zoomed range of parameters.} \label{fig04}
\end{figure*}

\section{Results: simulations vs observations}
\label{sec:Sec3}

Let us now discuss the numerical simulations that we want to compare
with observations in order to select the range of the potential
parameters (3.4). As we will see, analysis by hybrid gravity fixes
better the observational data  than the standard Keplerian analysis.

\subsection{Numerical calculation of S2 star orbit and
orbital precession}

The simulated orbits of S2 star around the central object in hybrid
gravity (blue solid line) and in Newtonian gravity (red dashed line)
for $\phi_0$ = -0.00033 and $m_\phi$ = -0.0028 (left panel), as well
as for $\phi_0$ = -0.000033 and $m_\phi$ = -0.00028 (right panel)
during 5 orbital periods, are presented in Fig. \ref{fig01}. As it
can be seen from this figure, hybrid gravity causes the orbital
precession in the same direction as GR, but precession angle is much
bigger. When both $\phi_0$ and $m_\phi$ are decreased for an order
of magnitude, the precession is much smaller (see the right panel of
Fig. 1). This analysis also shows that Keplerian orbit is recovered
when $\phi_0$ and $m_\phi$ tend to 0.

We calculate orbital precession in hybrid modified gravity potential
and results are reported in Fig. \ref{fig02} as a function of
$\phi_0$ and $m_\phi$. Assuming that the hybrid potential does not
differ significantly from Newtonian potential, we derive the
perturbed potential as

\begin{equation}
V(r) = \Phi \left( r \right) - {\Phi_N}\left( r
\right)\begin{array}{*{20}{c}}
;&{{\Phi_N}\left( r \right) =  - \dfrac{{GM}}{r}}
\end{array}.
\label{equ04}
\end{equation}

\noindent The obtained perturbing potential is of the form:

\begin{eqnarray}
V(r)=
\frac{G}{1+\phi_0}\left[1+\left(1/3\right)e^{-m_\phi
r}\right] M \phi_0/r.
\end{eqnarray}

\noindent and it can be used for calculating the precession angle
according to Eq. (30) in Ref. \cite{adki07}:

\begin{equation}
\Delta \theta = \dfrac{-2L}{GM e^2}\int\limits_{-1}^1 {\dfrac{z \cdot
dz}{\sqrt{1 - z^2}}\dfrac{dV\left( z \right)}{dz}},
\label{equ06}
\end{equation}

\noindent where $r$ is related to $z$ via: $r = \dfrac{L}{1 + ez}$.
By differentiating the perturbing potential $V(z)$ and substituting
its derivative and ($L = a\left( {1 - {e^2}} \right)$) in the above
Eq. (\ref{equ06}), and taking the same values for orbital elements
of S2 star like in Ref. \cite{bork12} we obtain numerically, for
$\phi_0$ = -0.00033 and $m_\phi$ = -0.0028, that the precession per
orbital period is $3^\circ.26$.

Graphical representation of precession per orbital period for 
$\phi_0$ in the range $[-0.0009, -0.0002]$ and $m_\phi$ in $[-0.0034,
-0.0025]$ is given in the left panel of Fig. \ref{fig02}. As one can
see, the pericenter advance (like in GR) is obtained. The precession
per orbital period for $\phi_0$ in the range $[-0.0004, -0.0002]$ and
$m_\phi$ in $[-0.0029, -0.0027]$ is given in the right panel of Fig.
\ref{fig02}.

\subsection{Comparison of theoretical results and observations}

Let us give some constraints on parameters $\phi_0$ and $m_{\phi}$
of hybrid gravity potentials according to current available
observations of S2 star orbit. However we should note that the
present astrometric limit is still not sufficient to definitely
confirm that the S2 orbit deviates from the Keplerian one, but there
is great probability that it is the case because the astrometric
accuracy is constantly improving from around 10 mas during the first
part of the observational period, currently reaching less than 0.3
mas (see reference \cite{frit10}). There are also some recent
studies that provide more and more evidence that the orbit of S2
star is not closing (see e.g. Fig. 2 in paper \cite{meye12}). In
this paper we fitted the NTT/VLT astrometric observations of S2
star, which contain a possible indication for orbital precession
around the massive compact object at Galactic Centre, in order to
constrain the parameters of hybrid gravity potential, since this
kind of potential has not been tested at these scales yet. We have
to stress that in the reference \cite{gill09b} on page 1092, Fig.
13, authors presented the Keplerian orbit, but in order to obtain
it they had to move the position of central point mass. In that way
they implicitly assumed orbital precession. In our orbit calculation
we do not need to move central point mass in order to get a
satisfactory fit.

In fact, our comparison with astronomical observations represents
upper bounds for precession angle on deviation from GR. The most
probably results for precession are in between these upper bounds
and GR results. In future, using more precise astronomical
observations, we could obtain more accurate results.

Figs. \ref{fig03} and \ref{fig04} present the maps of the reduced
$\chi^{2}$ over the $\phi_0 - m_\phi$ parameter space for all
simulated orbits of S2 star which give at least the same or better
fits than the Keplerian orbits. These maps are obtained by the same
fitting procedure as before. As it can be seen from Figs. 
\ref{fig03} and \ref{fig04}, the most probable value for the
parameter $\phi_0$ in the case of NTT/VLT observations of S2 star is
between -0.0009 and -0.0002 and for the parameter $m_{\phi}$ is
between -0.0034 and -0.0025 (see the darkest regions in Figs.
\ref{fig04}). In other words, we obtain reliable constraints on the
parameters $\phi_0$ and $m_{\phi}$ of hybrid modified gravity. The
absolute minimum of the reduced $\chi^{2}$ ($\chi^{2}=1.503$) is
obtained for $\phi_0$ = -0.00033 and $m_\phi$ = -0.0028,
respectively.

We simulated orbits of S2 star around the central object considering
both the hybrid gravitational potential and the Newtonian potential.
Our analysis shows that the hybrid modified gravity potential
induces the precession of S2 star orbit in the same direction of GR.
We used these simulated orbits to fit the observed orbits of S2
star. The best fit (according to NTT/VLT data) is obtained for the
$\phi_0$ from between -0.0009 and -0.0002, and for the $m_{\phi}$
between -0.0034 and -0.0025. This range corresponds to scale
parameter $m_{\phi}^{-1}$ from (1/0.0034) AU to (1/0.0025) AU
($\approx 300-400$ AU, i.e $1.4-1.9$ mpc) which is comparable to the
size of S2 star orbit.

We believe that comparison with astronomical observation is
important, and data we used are the best currently published and
available. GR predicts that the pericenter of S2 star should advance
by $0^\circ.18$ per orbital revolution \cite{gill09b}. Using our
fitting procedure, we get a much bigger precession $3^\circ.26$.
Figure 2 in this paper gives theoretically calculated precession per
orbital period for hybrid gravity of $\phi_0 - m_\phi$ parameter
space. In the future, with much more precise data maybe observation
will find smaller value of precession, and using Figure 2 (i.e. the
same procedure), we will be able to get again hybrid gravitational
parameters $m_\phi$ and $\phi_0$, hoping that observations will give
smaller values. We calculated the map of parameters theoretically for
broad range of precession angles. More precise observations probably
will change best fit parameters, but procedure for theoretical 
calculation will be the same.

\section{Conclusions}
\label{sec:Sec4}

In this paper, the orbit of S2 star around the galactic Centre has
been investigated in the framework of the hybrid modified gravity.
Using the observed positions of S2 star, we constrained the
parameters of hybrid modified gravity. Our simulation results are:

\begin{enumerate}
\item the range of values for $\phi_0$ parameter, coming from S2
star, is between -0.0009 and -0.0002;
\item the range of $m_{\phi}$ is between -0.0034 and -0.0025;
\item precession of S2 star orbit, in the hybrid modified gravity
potential, has the same direction as in GR, but the upper limit in
magnitude is much bigger than GR.
\end{enumerate}

The above results allow to compare the orbital motion of S2 
star in the framework of hybrid gravity with analogous results in 
other theories. In particular, hybrid gravity can be compared with  
metric $f(R)$ models, discussed in \cite{bork12,bork13} and with 
$f(R, \phi)$, discussed in \cite{capo14}. Also in these papers, the 
motion of S2 star has been studied according to the effective 
gravitational potentials achieved in the weak field limit. As 
discussed above, the main reason to introduce hybrid gravity lies on 
the fact that models like $f(R)$ gravity (both in metric and 
Palatini formalism) and $f(R, \phi)$ gravity suffer problems in   
passing the standard Solar System tests \cite{chib03,olmo05}. On the 
other hand, as reported in \cite{capo13b}, hybrid gravity allows to 
bypass shortcomings deriving from local tests and connect models to 
galactic dynamics and late time cosmic acceleration. Using  S2 star 
orbits, it is possible to achieve additional constraints at 
sub-parsec scales and promote this model with respect to other 
extended gravity approaches.

In particular, $\phi_0$ and $m_\phi$ are the specific parameters of 
hybrid gravity and differ from $f(R)$ gravity models both in metric 
and Palatini formalism. In the case of $f(R, \phi)$ gravity, it is 
possible to achieve a Sanders like potential; the parameter $m_\phi$ 
is also present and could have the same value, but the parameter 
$\phi_0$ of hybrid gravity and $\alpha$ of $f(R,\phi)$ differ  
\cite{capo14}. The two effective gravitational potentials, in the 
week field limit, have similar, but not the same forms at sub-parsec 
scales.

In conclusion, the comparison of the observed orbits of S2 star and
theoretical calculations performed by the hybrid modified gravity
model can provide a powerful method for the observational test of
the theory, and for observationally discriminating among the
different modified gravity models. It seems that hybrid gravity 
potential is sufficient in addressing the problem of dark matter at 
galactic scales \cite{capo13b}, and it gives indications that 
alternative theories of gravity could be viable in describing 
galactic dynamics.

Furthermore, orbital solutions derived from such a potential are
in good agreement with the reduced $\chi^2$ deduced for Keplerian
orbits. This fact allows to fix the range of variation of $\phi_0$
and $m_{\phi}$. The precession of S2 star orbit, obtained for the 
best fit parameter values ($\phi_0$ from -0.0009 to -0.0002 and 
$m_\phi$ from -0.0034 to -0.0025), has the positive direction, as
in GR, but for these values of parameters, we obtain much larger
orbital precession of S2 star in hybrid gravity compared to
prediction of GR.

We can conclude that hybrid gravity effective potential is probably 
the best candidate among the other considered gravity models such 
as e.g. $R^n$ \cite{bork12,zakh14}, Yukawa-like \cite{bork13} and 
Sanders-like \cite{capo14} to explain gravitational phenomena at 
different astronomical scales.

It is important to stress that our comparison with astronomical
observations represents only upper bounds for precession angle on
the deviation from GR. Although observational data seem to indicate
that the S2 star orbit is not Keplerian, the nowadays astrometric
limits are not sufficient to unambiguously confirm such a claim. We
hope that forthcoming observational data will allow more accurate
measurements of stellar positions.

A final remark is due now. From an astrophysical point of view, the
main motivation to introduce hybrid gravity is to address the
problems of dark matter and dark energy \cite{capo13b}. First of
all, we have to say, according to the observations, that dark matter
has very negligible effects around the Galactic Centre \cite{genz10}. 
Despite of this fact, here we adopted hybrid gravity dynamics only
to fit the orbit of S2 star around the Galactic Centre. The interest
of the reported results, if confirmed, lies on the fact that hybrid
dynamics is independent of the dark issues but can be connected to a
fine analysis of geodesic structure. In other words, the further
gravitational degrees of freedom, coming from hybrid gravity,
contribute to dynamics as soon as orbital analysis related to GR is
not sufficient to describe in detail peculiar situations as those
around the Galactic Centre. However, in order to better confirm this
statement, one needs more precise astronomical data describing
stellar dynamics around Galactic Center.

\paragraph{Acknowledgments}
D.B., P.J. and V.B.J. wish to acknowledge the support by the 
Ministry of Education, Science and Technological Development of the
Republic of Serbia through the project 176003. S.C. acknowledges the
support of INFN ({\it iniziative specifiche} QGSKY and TEONGRAV).
All authors acknowledge support by Bilateral cooperation between 
Serbia and Italy ''Testing Extended Theories of Gravity at different 
astrophysical scales'' and by ''NewCompStar'', COST Action MP1304. 
D.B. would like to thank to Dr. A.F. Zakharov for many usefull
discussions.

\end{document}